\documentclass[prb,twocolumn,amsmath,amssymb]{revtex4}
\begin{document}
\author{Kevin Leung$^*$ and Craig~J.~Medforth}
%Raid~Haddad,$^3$ and John.~A.~Shelnutt$^2$}
\affiliation{Sandia National Laboratories, MS 1415 \& 1349 Albuquerque,
NM 87185 \protect\\
$^*$Email: {\tt kleung@sandia.gov}
}
%\date{\today}
\title{{\it Ab initio} molecular dynamics study of manganese porphine
hydration and interaction with nitric oxide}

\input epsf
%\ssp

\begin{abstract}
 
We use {\it ab initio} molecular dynamics (AIMD) and the DFT+U method
to compute the hydration environment of the manganese ion in
manganese (II) and manganese (III) porphines (MnP) dispersed in liquid water.
These are intended as simple models for more complex water soluble porphyrins,
which have important physiological and electrochemical applications.  The
manganese ion in Mn(II)P exhibits significant out-of-porphine
plane displacement, and binds strongly to a single H$_2$O molecule in
liquid water.  The Mn in Mn(III)P is on average coplanar with the porphine
plane, and forms a stable complex with two H$_2$O molecules.  The residence
times of these water molecules exceed 15~ps.  The DFT+U method correctly
predicts that water displaces NO from Mn(III)P-NO, but yields an
ambiguous spin state for the MnP(II)-NO complex.
 
\end{abstract}
 
\maketitle
 
\section{Introduction}
 
Porphyrins dispersed in water, coated on electrodes, and self-assembled
into nanotubes have important applications as
sensors\cite{handbook,nitric1,nitric2,nitric3,sol1,sol2} and light-induced
water-splitting and hydrogen production.\cite{water}  Simple
manganese porphyrins have been used as readily isolated experimental
models for studying isoelectronic Fe(III) systems.\cite{nitric_rev1}
Water soluble Mn(II) and Mn(III) porphyrins are
particularly useful for detection of nitric oxide
in cell tissues, and for distinguishing
them from nitroxyl (NO$^-$ and HNO) compounds.\cite{nitric1,nitric2}
The relaxation processes of water soluble manganese and iron
{\it tetra-p}-sulfonatephenyl porphyrins (TSPP) in NMR-compatible time
scales have also been the subject of significant experimental
studies.\cite{expt1,expt2,expt3,expt4}
 
Despite their importance and widespread appearance in
technological and biological settings, theoretical studies of
transition metal porphyrins in aqueous media have been rare.  This
is in part due to the difficulty of modeling transition metal systems
in general.  In liquid water, bare divalent and trivalent first row
transition metal ions are octahedrally 6-coordinated due to their
partially filled 3d electron shells.  While density functional
theory (DFT) successfully predicts this feature,\cite{dft_water}
sophisticated classical force fields
with 3-body terms or quantum mechanics/molecular mechanics methods
are needed to reproduce such hydration structures.\cite{threebody,rode}
Transition metal porphyrins present an inherently interesting case.\
The molecular framework has an overall -2$e$ charge resulting from the
nitrogen atoms chelating the metal ions.  How these chelated ions in
porphyrins interact with water will be affected by the ion size, spin
state, and porphyrin conformations.
 
Predicting the spin state of transition metal porphyrins 
presents a considerable theoretical challenge.  While DFT
is formally exact, practical implementations depend on the
choice of the approximate exchange correlation functional, not all of which
yield the correct spin state in transition metal complexes.\cite{reiher} This
is particularly true for first row transition metal porphyrins.\cite{ghosh}
It has been shown that the hybrid
functionals like B3LYP\cite{b3lyp} yields the correct stable high spin
ground state for manganese (II) porphine (Mn(II)P) and manganese
(III) porphine (Mn(III)P),\cite{leung}
while non-hybrid functionals like PBE,\cite{pbe} lacking
a long range Hartree-Fock exchange component, predict the incorrect
intermediate spin state for Mn(II)P.  If constrained to the high spin
state, PBE still yields a poor Mn(II)P molecular geometry.\cite{leung,dimer}
However, hybrid functionals are far more expensive to apply than non-hybrids
when applying periodic boundary conditions appropriate to condensed
phase systems such as liquid water or metal surfaces.
This problem has been circumvented~\cite{leung} by applying the DFT+U
method on the Mn 3d electrons.\cite{ldau}  The screened coulomb
(``$U$'') term in DFT+U increases repulsion between electrons
in low-spin, partially filled $d$-electron systems.  In the approach
used in 
%Ref.~\cite{leung},
Ref.~19,
this term was parameterized using B3LYP spin splittings, and then
the technique was applied to model MnP adsorbed on gold electrodes.
The DFT+U method has also proved accurate for Fe-based catalysts.\cite{marz}
 
In this paper, we extend this previous work by conducting DFT+U based
{\it ab initio} molecular dynamics (AIMD) simulations of Mn(II) and
Mn(III) porphines in liquid water.  Porphines 
not substituted with ionic side groups are mostly water-insoluble;
our MnP molecules are intended as simple
models for soluble porphyrins such as (MnTSPP),\cite{expt1,expt2,expt3}
which are too large and computationally costly to simulate using
AIMD.\cite{note0}
The elucidation of the hydration environment of the metal ion in
porphyrins is a prerequesite for studying the binding of additional
ligands to the metal site of water-soluble porphyrins.  AIMD hydration
studies are potentially useful for modeling electrochemical half-reactions
involving porphyrins; such calculations have already been
performed for transition metal ions solvated in water.\cite{sprik}  Our work
also suggests that the metal ion hydration structure assumed in
the analysis of NMR relaxation experiments on
Mn(II)TSPP may need to be re-examined.\cite{expt1,expt2}  
 
Finally, we briefly consider the binding between Mn(II)P and Mn(III)P and
nitric oxide (NO) in the gas phase and in water.  Such complexes
are classified according to the format \{MNO\}$^n$,\cite{nitric_rev}
where $n$ is the number
of metal $d$-electrons plus the unpaired electron contributed by NO.
For $n \leq 6$, the M-NO bond should be short ($\sim 1.65$~\AA),
and the M-N-O angle should be $\sim 180^o$.\cite{nitric_rev}
Mn(II) porphyrins are $n=6$ complexes, and X-ray structures
of Mn(II) porphyrins ligated to NO indeed exhibit
bond lengths and angles appropriate to
\{MNO\}$^6$.\cite{nitric_rev1,nitric_rev2}  To our knowledge,
no X-ray structures of NO-ligated Mn(III) porphyrins have been reported.
Water soluble Mn(III) porphyrins are found not to
complex with NO in aqueous conditions.\cite{nitric3} As a result,
electrochemical switching
of the Mn oxidation state is potentially useful for detecting NO
in cell tissues, and in distinguishing
NO from HNO, which binds to Mn porphyrins in both oxidation states.
 
Ideally, the same DFT method and/or functional can be used to predict
all properties (energetics, structures, spectral properties) in all aqueous
phase applications.
There have been several reported successes of DFT treatment of
Fe porphyrin-NO complexes.\cite{feno,feno1}  These studies have
focused on the molecular structures and vibrational properties,
and not the binding energies between porphyrins and  nitric oxide.
In this work, we
critically examine the effect of applying different exchange-correlation
functionals on the spin state, molecular geometry, and binding energies
of MnP-NO.  We find that the DFT+U method successfully
predicts that the Mn(III)P-NO complex dissociates in liquid water environments.
However, we also conclude that neither DFT+U nor existing, widely used
DFT functionals enjoy universal succcess in modeling nitric oxide
ligation.  This suggests that substantial functional development and
refinement are necessary.
 
This paper is organized as follows.  Section 2 describes the method
used.  Section 3 discusses the structure of water around the manganese
ions in porphines.
%Porphine conformational changes are described in Sec.~4,
MnP-NO binding is briefly described in Sec.~4, 
and Sec.~5 concludes the paper with further discussion.
 
\section{Methods}
 
We conduct {\it ab initio} molecular dynamics (AIMD) based on the
DFT+U\cite{ldau} method implemented\cite{vaspldau} in the VASP
code version 4.6.\cite{vasp}  Our VASP calculations apply
the PBE exchange correlation functional, the projected-augmented
wave (PAW) method\cite{paw} and the standard VASP
suite of PAW pseudopotentials.  These include the Mn pseudopotential
with pseudovalent $2p$ electrons, and the oxygen,
carbon, nitrogen, and hydrogen pseudopotentials with default
energy cutoffs of 400, 400, 400, and 250~eV, respectively.

The Mn 3d orbitals are augmented with screened column
(``$U$'') and exchange (``$J$'') terms of $U=4.2$ and $J=1.0$~eV,
respectively.  These values have been shown to reproduce
Mn(II)P high-spin/intermediate spin splittings predicted by
the B3LYP functional,\cite{leung} and predict the high spin
Mn(III) porphyrin spin state observed in experiments.  For more details, see 
%Ref.~\cite{leung}
Ref.~19
and the supporting information therein.
 
%revised

\begin{figure}
\centerline{ \hbox{\epsfxsize=3.30in \epsfbox{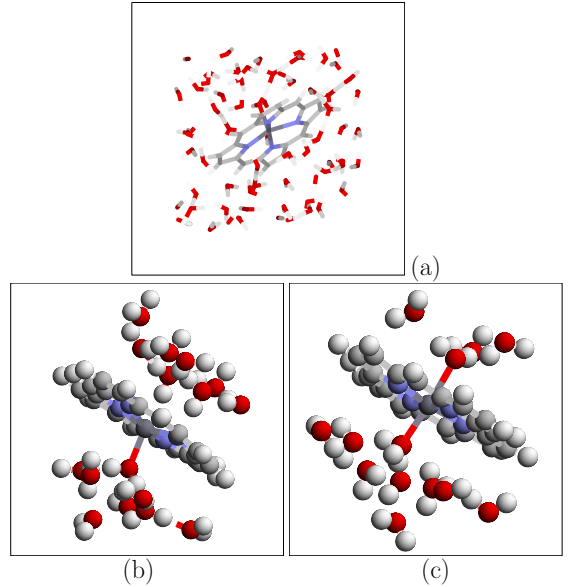}}}
\caption[]
{\label{fig1}
Representative snapshots of (a) \& (b) Mn(II)P in water; (c)
Mn(III)P in water.
Except in panel (a), only the water molecules close to the Mn ion are
depicted. }
\end{figure}

A plane wave energy cut-off of 400~eV and a wavefunction convergent criterion
of 10$^{-5}$~eV at each of the 0.5~ps time step in the Born-Oppenheimer
AIMD simulations.  This work focuses on structural properties; hence,
deuterium is substituted for all protons so that a larger time step
can be used.  A Nose thermostat with an estimated time constant of 
20~fs maintains the average temperature at T=375~K.  
With these parameters, the drift in total energy is 0.9~Kelvin
per picosecond (i.e., $\sim 0.004$ mHt/atom/ps) or less.  This small
energy drift is absorbed by the thermostat.  Since the underlying PBE
exchange correlation functional yields overstructured water and slow
dynamics at T=300~K,\cite{overstructure} the elevated temperature is
necessary to obtain reasonable water structure and diffusion timescales.
(See discussions in the next section.)
These factors imply that the trajectories do not capture the real-time
dynamics.  Instead, the time dependence observed
is used to confirm that the system has reached equilibrium, and
it provides a time scale for thermal fluctuations under these
simulation conditions.  Trajectory lengths of at least 15~ps
are used to collect statistics after equilibration runs have stabilized
the potential energy.
 
The simulation cell is cubic with a linear dimension of 13.65~\AA.  The
planar porphine molecule is inserted diagonally into the cell.  This
ensures that there are 5 to 6 layers of water molecules separating
the periodically replicated MnP in the direction perpendicular to
the porphine ring (Fig.~1).  The water density in
the cell is determined using a grand canonical
Monte Carlo (GCMC) simulation, using the Towhee code,\cite{towhee}
SPC/E rigid water molecule force fields,\cite{spce} and porphine force
fields appropriate for NiP,\cite{shelforce} which, with the exception
of the metal ion parameters, should be reasonable
for porphine atom-H$_2$O interactions.  The Mn(II)P is held
frozen in place in the GCMC simulations.
It is found that 70 water molecules reside in the simulation cell.
In the absence of Mn(II)P, the 13.65~\AA$^3$ cubic
cell would contain 85 water molecules at 1.0 g/cc density.
 
The water content for Mn(II)P-NO is similarly determined using GCMC,
resulting in 68 H$_2$O molecules in the simulation cell.  The nitric oxide
force field used is similar to that of 
%Ref.~\cite{no}.  
Ref.~37.
NO is essentially a hydrophobic molecule that interacts weakly with water.
 
%The porphine ring conformational changes are analyzed using the
%Normal-Coordinate Structural Decomposition (NSD) method developed
%by John Shelnutt, Craig Medforth, and coworkers.\cite{craig}

All liquid phase pair correlation functions $g(r)$
are computed with a 0.1~\AA\, bin size, except that a 0.05~\AA\,
bin size is used for pure liquid water.
 
B3LYP hybrid functional\cite{b3lyp} calculations of 
gas phase porphine complexes are performed using the program
Gaussian 03.\cite{g03}  We first optimize the geometries using
the LANL2DZ basis set.  Then we switch to the more accurate
6-311+G(d,p) basis, and perform geometric optimization for 3~to~5
steps until the magnitude of each Cartesian component of
the force on each atom is below $\sim 0.05$~eV/\AA.
(VASP gas phase calculations utilize a similar criterion for
attaining optimal geometry.)  We find that
the PBE Mn(II)P-H$_2$O binding energy
computed using the 6-311+G(d,p) basis set, and that computed using
the VASP code with plane wave basis sets, agree to within 
24~meV.
 
\section{Hydration structures of manganese porphines}
 
First we consider the hydration structure of Mn(II)P in water.
Simple, electrostatic-plus-Lennard Jones interactions between water
and the porphine molecule are applied during the GCMC pre-equilibration
simulation.  These force fields predict that a H$_2$O ligates to
the metal ion on each side of the porphine plane, but the overall metal
coordination structure is strongly distorted from the ideal octahedral
geometry.  Applying AIMD on this classical force field-generated starting
configuration leads to substantial structural changes.  We find
that a 8~ps AIMD trajectory is needed to equilibrate the potential
energy of the system.  AIMD simulation of Mn(III)P
in water is started from an equilibrated Mn(II)P configuration, and
the potential energy of the system also converges to a plateau value
within 8~ps.  These long equilibration times help ensure that
the statistics collected in the subsequent 15~ps production run are
independent of initial conditions.
 
Figure~1a depicts a equilibrated hydration structure around
Mn(II)P and confirms that there are several layers
of water molecules buffering the Mn ion from its periodic image.
This suggests that the hydration
environment of the Mn ion adequately represents bulk water
boundary conditions.  
%The edges of the planar MnP molecule are separated
%by only 2-3 H$_2$O layers from their periodic images.  However, the
%hydration of the edges of the unsubstituted porphine is of less
%interest because MnP is likely insoluble in water.  Instead,
%MnP is taken as a model for water-soluble MnTSPP, which
%is too large to simulate using AIMD methods.
 
The representative configuration depicted in Fig.~1b focuses
on H$_2$O close to the Mn ion.  Unlike the classical force field results,
the Mn ion in Mn(II)P strongly binds to one rather than two
H$_2$O molecules.  As discussed below, this can be related to the significant
out of porphine plane motion exhibited by the Mn(II) ion.
Figure~1c depicts Mn(III)P in water.  Mn(III)P is on average
coplanar with the porphine plane, and the Mn binds to two water molecules.
These snapshots encapsulate this section's main results, which will
be analyzed in some detail below.
 
\begin{figure}
\centerline{ \hbox{\epsfxsize=2.70in \epsfbox{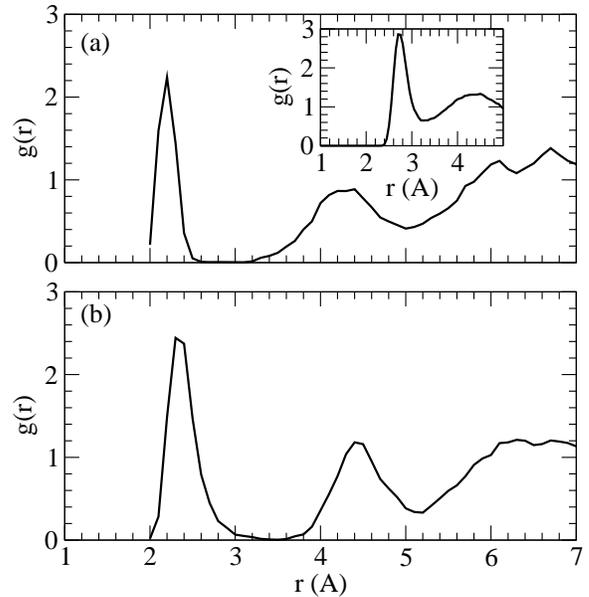}}}
\caption[]
{\label{fig2}
Pair correlation functions $g(r)$ between the Mn ion and
the oxygen sites of H$_2$O for (a) Mn(II)P; (b) Mn(III)P.
Inset: $g_{\rm OO}(r)$ for liquid water.}
\end{figure}
 
The Mn(II)-H$_2$O pair correlation function $g_{\rm Mn-O_w}(r)$ is plotted in
Fig.~2a, and it confirms the structure shown in Fig.~1b.
The first peak in the Mn-O$_{\rm w}$ $g(r)$ integrates to 1.0
from $r=0$ to $r=3.0$~\AA, which is the first $g(r)$ minimum.  The sharp
peak and deep minimum clearly show that this water molecule is strongly bound
to the Mn(II) ion.  They are reminiscent of divalent cation $g(r)$ in
liquid water.\cite{rode,bery}  In fact, over the entire 15~ps production
run trajectory, the same water molecule remains bound to the Mn ion,
although during the 8~ps equilibration trajectory, exchange of Mn first
hydration shell water molecules with the bulk water region has occurred.
Experimentally, the time scale for proton exchange that involves
H$_2$O bound to Mn(II) is known to be on the order $10^{-8}$ to
$10^{-7}$~s.\cite{expt_time}

The inset to Fig.~2 depicts the pure water $g_{\rm OO}(r)$,
computed using the PBE functional, 32 water molecules at 1.00~g/cc
light water density, and with the thermostat set at T=375~K.  
It confirms that PBE water at T=375~K exhibits less overstructuring
compared to PBE water at T=300~K.  For example, the $g_{\rm OO}(r)$
first peak exhibits a peak height of 2.9 density units at T=375~K, which
is in reasonable agreement with the value of 3.1 reported by Sit
{\it et al.}.\cite{marz1}  The small discrepancy can be attributed to
statistical uncertainties and/or the use of different bin sizes.
In contrast, at T=300~K, this peak height exceeds 3.4 density
units,\cite{galli2} which is considerable larger than the experimental
value.\cite{marz1}  Our choice of T=375~K is thus reasonable
for MnP hydration studies.  While we have imposed a water density estimated
using classical force fields and the GCMC method, we note that ion
hydration structure appears insensitive to significant variations
($\sim 7$\%) in water density.\cite{rigid}  

The stability of the single-H$_2$O coordination structure of Mn(II)P
is closely related to the significant Mn out-of-plane motion.
Figure~3a depicts the evolution of this displacement along
the AIMD trajectory.  
%Unlike in static gas phase calculations, the
%porphine ring carbon atoms exhibit considerable distortions from
%a planar geometry.  
Here, at each time step, we perform a
least square-fit to generate a plane, $ax+by+cx=1$, through the 4~nitrogen
atoms, and $\delta_{\rm Mn}$ is obtained as the smallest distance between
the Mn atom and this plane.  
On average, $\delta_{\rm Mn}$ is 0.48~\AA\, for
Mn(II)P dispersed in water.

This large displacement can be related to the extremely flat
out-of-plane Mn potential energy surface predicted by the DFT+U 
method for high spin Mn(II)P in the gas phase (Fig.~3b).
Thus, upon binding to a ligand like a H$_2$O, the Mn ion is readily
displaced out of the porphine plane.  We find that this displacement is
already 0.225~\AA\, for the Mn(II)P-H$_2$O complex in the gas phase, and
this geometric feature explains the inability of the Mn ion to bind to a
second H$_2$O in a liquid water enviroment.  Note that the out-of-plane
displacement of the Mn ion is accompanied by only a small amount out-of-plane
deformation of the porphine ring (approximately
0.1~\AA\, total deformation as determined by Normal Coordinate Structure
Decomposition; see Table S1).
 
The out-of-plane Mn motion exhibits a flat potential energy surface
because the high spin Mn(II) ion has a large effective
size. The Mn-N distance in high spin Mn(II)P
is predicted to be 2.09~\AA, in good agreement
with experiments.\cite{mnst_expt1,mnst_expt2}  Thus, energetically,
it is relatively unfavorable for Mn to fit in the center of
the ion-chelating site of the porphine molecule.
In contrast, the Mn(III) ion is smaller, with an Mn-N distance
of 2.045~\AA\, in MnClP,\cite{leung} and in isolated Mn(III)P
strongly prefers to be coplanar with the porphine plane
(Fig.~3c).  In liquid water, this preference persists
on average (Fig.~3a).  This geometry
allows effective binding to two ligands.  Indeed the sharply peaked
Mn-water first peak in $g(r)$ integrates to 2.0, indicating two
H$_2$O are strongly ligated to the Mn ion.  We do not observe exchange of
the two H$_2$O molecules bound to the Mn(III) ion with outer shell water
during the 15~ps trajectory.

A search of the Cambridge Crystallographic Data Center (CCDC)
yields 15 crystal structures of Mn porphyrins with one or more
coordinated water molecules.  The manganese ion exhibits the
+3 oxidation state in all of them.  The experimentally determined
Mn-O distances in the Mn(III) porphyrins ($\sim 2.2$~\AA) correspond
closely to the first peaks in the pair correlation functions in
Fig.~2.  The average calculated Mn-N(porphine) distance
in the high-spin ($m=2$) Mn(III)P(H$_2$O)$_2$ energy minimized
using the DFT+U is 2.028~\AA.  This again agrees well with
the Mn-N(porphyrin) distances observed in Mn(III) porphyrin
complexes with two axial water ligands
(e.g., Mn(III)TPP(H$_2$O)$_2$(BPh)$_4$, where the Mn-N distance
is 2.011~\AA).\cite{turner}   The porphine ring again shows very
little non-planar distortion in either the DFT+U gas phase
structure of Mn(III)P(H$_2$O)$_2$ (Table S2) or the crystal
structure of Mn(III)TPP(H$_2$O)$_2$(Bph)$_4$ (Table S3).

Gas phase B3LYP and DFT+U calculations of the MnP-(H$_2$O)$_2$
complex (Fig.~4, Table~\ref{table1}) provide more compelling
evidence that Mn in Mn(II)P coordinates to only one water molecule.
Both these methods predict that
there are two nearly isoenergetic configurations
for the high spin Mn(II)P-(H$_2$O)$_2$ cluster (Fig.~4).
The bidentate Mn(II) configuration, with the Mn ion residing
in the porphine plane (Fig.~4a), is only slightly more stable than
the one where Mn(II) is $\sim 0.4$~\AA\, out-of-plane and 
one H$_2$O is bound to Mn and the other coordinated to two
nitrogen atoms (Fig.~4b).  DFT+U 
predicts that the bidentate configuration is more stable
by only 15~meV, while B3LYP predicts that they are equally stable.
In contrast, with the system in either the intermediate or
the high spin state, PBE predicts that the structure depicted in
Fig.~4b is {\it unstable} and spontaneously relaxes
to that shown in Fig.~4a.  This emphasizes that there are
fundamental differences between DFT+U and PBE predictions
for gas phase properties.  These differences carry over into 
liquid phase AIMD simulations (Fig.~1a) favoring
configurations similar to Fig.~4b.\cite{note}
 
\begin{figure}
\centerline{ \hbox{\epsfxsize=2.70in \epsfbox{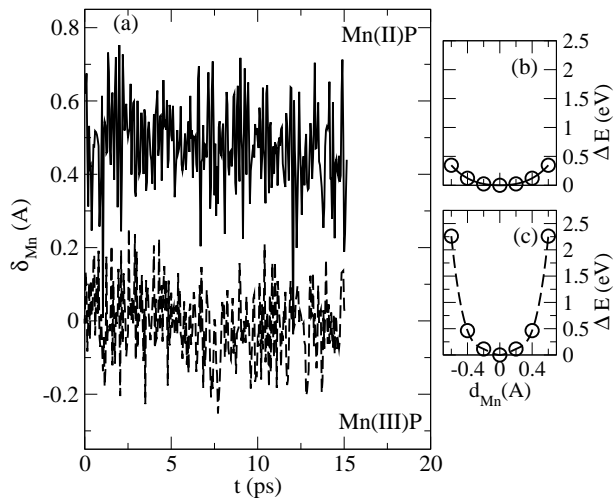}}}
\caption[]
{\label{fig3}
(a) Out-of-porphine plane displacement of Mn ion,
$\delta_{\rm Mn}$, over the AIMD trajectories for Mn(II)P and Mn(III)P.
(b) \& (c) Energies as functions of Mn out-of-plane displacement in the
gas phase for Mn(II)P and Mn(III)P, respectively
The porphine ring is held fixed as the Mn is displaced.}
\end{figure}

Because of the slow exchange rate of Mn first hydration shell
H$_2$O, it is important to check whether the Mn(II)P hydration
structure we observe depends on the initial configurations.
Thus, we start with a equilibrated Mn(III)P configuration, with
two water molecules ligated to the Mn ion,
and initiate a Mn(II)P AIMD simulation by adding one more
electron to the simulation cell.  Within 1~ps,
one of the first hydration H$_2$O detaches itself, the
Mn ion exhibits strong out-of-plane displacement, and the single
H$_2$O hydrated, 5-coordinated Mn(II) configuration is recovered.
Thus, this Mn(II)P hydration structure is robust.
 
The predicted hydration structure of Mn(II)P might possibly
resolve anomalies associated with paramagnetic
relaxation behavior of Mn(II)TPPS in water.\cite{expt2,expt3}
One model proposed in the NMR literature assumes a 6-coordinated
geometry, where 2 axial H$_2$O molecules bind
to the manganese ion in Mn(II)TPPS.\cite{expt3} In contrast,
AIMD predicts only one H$_2$O in the hydration shell and a breaking
of the inversion symmetry about the porphine plane, which now
exhibits a C$_{\rm 4v}$ symmetry.  This suggests that the underlying
assumptions of 
%Ref.~\cite{expt3}
Ref.~11
regarding coordination to water may need to be re-examined.  While
paramagnetic relaxation behavior is complex,
we propose that using a model with the Mn ion ligated to the AIMD
predicted number of water molecules might improve agreement between
experiments and the spin-Hamiltonian results used
to inteprete NMR data.\cite{expt2,expt3}
 
\begin{figure}
\centerline{ \hbox{\epsfxsize=2.70in \epsfbox{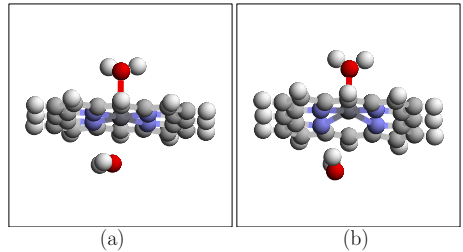}}}
\caption[]
{\label{fig4}
Two configurations of Mn(II)P-(H$_2$O)$_2$.  DFT+U and B3LYp
both predict that they are similar in energy, while PBE
predicts that the structure in (b) is unstable.}
\end{figure}

Figure~5 examines the
hydration environment of the four nitrogen atoms surrounding 
the Mn ion.  Integrating the $g_{\rm N-O_w}(r)$ to
its first minimum (3.8 and 3.9~\AA\, for Mn(II)P and Mn(III)P
respectively), we find that there are 1.76 and 2.83 H$_2$O
in the hydration shell of each nitrogen atom.
These spatial correlations partly originate from the one
(two) H$_2$O strongly coordinated to the Mn(II) (Mn(III)) ion in the
vicinity of the  nitrogen atoms, and not from water-nitrogen hydrogen
bonding.  A N-H hydrogen bond typically exhibits a
$g_{\rm N-H_w}(r)$ peak at $r \sim 1.8$~\AA, and a first minimum
at $r \sim 2.5$~\AA.\cite{hydrox}  Here, $g_{\rm N-H_w}(r)$ only
exhibits small shoulders.  Integrating $g_{\rm N-H_w}(r)$
to $r=2.5$~\AA\, yields only 0.10 and 0.02 H$_2$O in the first
coordination shell per
nitrogen atom for Mn(II)P and Mn(III)P, respectively.  We conclude
that water interacts weakly with the nitrogen atoms.

Note that our AIMD trajectories impose a high spin configuration,
known to be favorable in the gas phase.\cite{leung}
We have tested snapshots of the Mn(II)P trajectory to see if Mn(II)P
in water favors spin states at variance from those
predicted for the gas phase ground state.  We invariably observe that the
high spin configuration is favorable in these DFT+U generated AIMD snapshots
when using the DFT+U technique, in agreement with experiments
on MnTSPP in water.\cite{expt2,expt3}  The intermediate spin state is
favored in these snapshots when the PBE functional is used.  Thus,
dispersing Mn(II)P into liquid water does not alter the most stable spin
state; for each functional used, the gas phase and aqueous phase spin
state predictions are in agreement.  This is expected because water is not
a strong ligand.
 
\begin{table}\centering
\begin{tabular}{  c r c r r r r  } \hline
Species & Ligand & $m$ & PBE & B3LYP & DFT+U \\ \hline
%%Mn(II)P & NO & 0  &   -2.031 &  -0.545 &  -0.607 \\
Mn(II)P & NO & 0  &   -2.031 &  -0.524 &  -0.607 \\
Mn(II)P & NO & 1  &   -1.687 &  -0.728  &  -0.930 \\
Mn(II)P & NO & 2  &   NA & -0.022  &  -0.407 \\ \hline
Mn(III)P & NO & 1/2  &  -1.302 &  0.000$^*$ & +0.030$^*$ \\
%%Mn(III)P & NO & 3/2  &  - 0.651 &  -0.331 &  -0.401 \\
Mn(III)P & NO & 3/2  &  - 0.651 &  -0.322 &  -0.401 \\
Mn(III)P & NO & 5/2  &  -0.103 &  -0.077 &  -0.163 \\  \hline
Mn(II)P & H$_2$O & 5/2  &  NA &  -0.359 &  -0.345 \\ 
Mn(II)P & H$_2$O & 3/2  &  -0.283 &  NA &  NA \\  \hline
Mn(II)P & 2 H$_2$O & 5/2  &  NA &  -0.487 &  -0.516 \\ 
Mn(II)P & 2 H$_2$O & 3/2  &  -0.457 &  NA &  NA \\  \hline
Mn(II)P & NO + H$_2$O & 1 & NA  &  NA  &  -0.886 \\ 
Mn(II)P & NO + H$_2$O & 0  & -2.309  & NA  &  NA \\  \hline
Mn(III)P & H$_2$O & 2  &  -0.571 &  -0.629 &  -0.581 \\ \hline
Mn(III)P & 2 H$_2$O & 2  &  -0.997 & -1.045  &  -0.946 \\ \hline
Mn(III)P & NO + H$_2$O & 3/2 &  NA & NA  &  -0.590 \\ 
Mn(III)P & NO + H$_2$O & 1/2 &  -1.854 & NA  &  NA \\ \hline
\end{tabular}
\caption[]
{\label{table1} \noindent
Binding energies between MnP and NO or H$_2$O, in eV.\cite{zeropt}  PBE
and DFT+U results utilizes planar wave basis while B3LYP calculations are
performed using the 6-311+(d,p) basis.  $m$ is the total magnetic
polarization in the complex.  In some cases only the results for the
most stable spin state predicted for each DFT functional are listed.
$^*$Negligible binding energy predicted.
}
\end{table}
 
\section{Interactions with nitric oxide}
 
In this section, we examine the interaction of Mn(II)P and
Mn(III)P with NO.  Ideally, the DFT+U method, which predicts
the correct $s=5/2$ spin state for Mn(II)P, can be used
to treat MnP-NO binding in aqueous environments.  We will
show that DFT+U appears to yield reasonable binding energies.
Neither DFT+U nor B3LYP necessarily predicts the correct
spin states for the MnP-NO complexes.  However, since the energies
of different spin states are so similar, our conclusions
concerning binding energies appear to be robust.

We first consider gas phase predictions.
Table~\ref{table1} depicts the PBE, B3LYP, and DFT+U results based on
the PBE functional and $U=4.2$~eV, $J=1.0$~eV.
Here we use the notation
$m$ to denote the magnetic moment of the entire complex in the
simulation cell.
%instead of $s$ which is more appropriate for
%the orbital magnetic moment of a single paramagnetic species.
 
\subsection{Mn(II)P-NO}
 
PBE correctly predicts a low spin ground state for Mn(II)P-NO, with
a 1.617~\AA\, Mn-N (nitric oxide) bond length, and a
Mn-N-O angle of 173.6$^o$.  The Mn ion is displaced 0.342~\AA\, out of the
porphine plane, and the porphine ring shows little evidence ofo non-planar
deformation ($\sim 0.1$\AA, see Table S4).  These predictions are in good
agreement with the X-ray structure for nitrosyl(5,10,15,20-tetratolylporphinato)
manganese (Mn(TTP)NO), which reveals a 1.641~\AA\, M-NO bond length,
a 177.8$^o$ M-N-O angle, and a 0.337~\AA\, Mn out-of-plane
displacement.\cite{nitric_rev,nitric_rev1,nitric_rev2} 
The four other Mn(II)tetraarylporphyrins in the CCDC are 6-coordinated
but show similar Mn-NO distances (1.645-1.680~\AA).
The average Mn-N(porphyrin) distance in the PBE structure of Mn(II)P-NO
is 2.022~\AA, which compares favorably with the crystal structure
distance of 2.004~\AA.\cite{nitric_rev}  The crystal structure of
Mn(II)TPP-NO also shows a modest amount of non-planar deformation
($\sim 1.2$~\AA, Table S5) which may in part explain the slightly shorter
Mn-N (porphyrin distance.  PBE predicts a large (2.03~eV) MnP-NO binding
energy (Table~\ref{table1}).

\begin{figure}
\centerline{ \hbox{\epsfxsize=2.70in \epsfbox{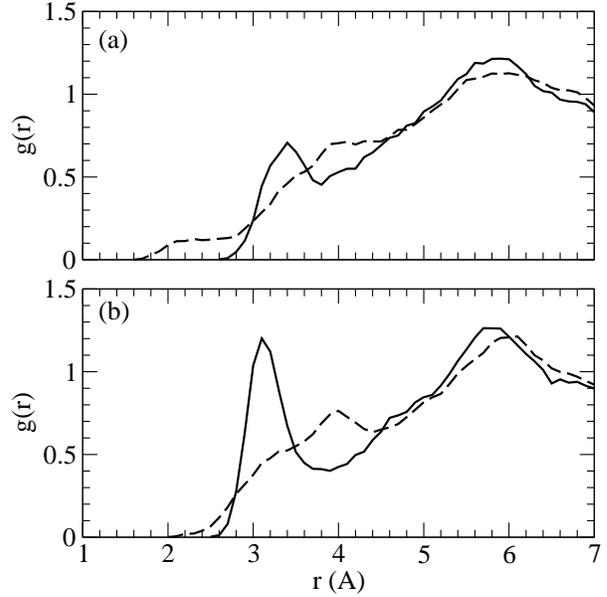}}}
\caption[]
{\label{fig5}
Pair correlation functions, $g_{\rm N-O_w}(r)$ (solid lines)
and $g_{\rm N-H_w}(r)$ (dashed lines), between the nitrogen atoms and
the oxygen/hydrogen sites of H$_2$O for (a) Mn(II)P; (b) Mn(III)P.}
\end{figure}
 
DFT+U and B3LYP results are qualitatively and even quantiatively similar
to each other but disagree with the experimental data.  They predict,
incorrectly, that the $m=1$,
intermediate spin configuration is more stable than the low spin state,
by 0.20 and 0.32~eV, respectively.  
For this ground state, the Mn-N$_{\rm NO}$
distance is 1.91 (1.89)~\AA\, the Mn-N-O angle is 142.2 ( 144.2)$^o$, the
out-of-plane Mn displacement is 0.277 (0.323)~\AA,
and the MnP-NO binding energy is 0.93 (0.73)~eV.
DFT+U further predicts that the low-spin configuration
still exhibits a considerable binding energy of 0.61~eV, with a 1.766~\AA\,
Mn-NO bond length and 174$^o$ M-N-O bond angle, respectively.
 
We note that Gaussian B3LYP calculations reveal some
spin-contamination for the $m=1$ intermediate spin state.
The VASP package does not calculate spin contaminations, but it
is likely that VASP-based DFT+U calculations also contain a similar
degree of spin-contamination for this spin state.  We emphasize that
only the NO-ligated species exhibit such a contamination.  Since both
the energies of the $m=0$ and $m=1$ spin states agree to within 0.3~eV,
the order of magnitude of our binding energies will likely not be 
strongly affected by the ambiguity regarding the spin state.
 
\subsection{Mn(III)P-NO}
 
To our knowledge, no X-ray structures for NO-ligated Mn(III) porphyrins
exist.  Such hypothetical systems would be \{M-NO\}$^5$ complexes, which
in principle should exhibit a low-spin ($m=1/2$), short
Mn-N$_{\rm NO}$, linear Mn-N-O geometry.\cite{nitric_rev}  PBE
predicts such a low spin structure, but also yields an anomalously
large MnP-NO binding energy of 1.30~eV (Table~\ref{table1}; see below).
This structure has an out-of-plane Mn displacement of 0.36~\AA, which
is comparable to that seen for low spin Mn(II)P-NO.
 
Both DFT+U and B3LYP predict negligible binding energies for NO with
low spin Mn(III)P.  The $m=3/2$ intermediate spin configuration, on the
other hand, yields a small binding energy of 0.40 (0.32)~eV and a small
out-of-plane Mn displacement of 0.16 (0.10)~\AA.  While no experimental
data for binding energy is available, DFT+U predicts
that the Mn(III)P-H$_2$O complex is more favorable than the Mn(III)P-NO
complex, while the reverse is true for PBE.  This suggests that
H$_2$O will displace NO in a DFT+U based AIMD simulation, consistent
with experiments on water-soluble Mn(III)
porphyrins.\cite{nitric1,nitric2,nitric3}  In contrast,
because of the strong binding energy with NO, AIMD simulations based on the
PBE exchange correlation functional will likely not lead to Mn(III)P-NO
dissociation in water.\cite{note1}

Gaussian calculations again reveal some spin-contamination for both
the $m=1/2$ and $m=3/2$ states.  In contrast, no spin-contamination is observed
for either Mn(II)P or Mn(III)P-Cl.\cite{leung}  This underscores that
MnP-NO systems are difficult to model using DFT methods.
 
\subsection{MnP-NO in water}
 
Next, we investigate the stability of MnP-NO complexes in liquid water by
conducting AIMD simulations.   We pre-equilibrate
hydrated Mn(II)P-NO configurations from a GCMC
run as before, holding MnP-NO rigid in the DFT+U optimized,
intermediate spin configuration, and then conduct AIMD simulations
for 3~ps.  Unlike for Mn(II)P ligated to H$_2$O, the potential energy
reaches a plateau on a sub-picosecond timescale.  We find that the
Mn ion in Mn(II)P-NO complex is stable in water, and does not 
ligate to any additional H$_2$O molecules.
 
\begin{figure}
\centerline{ \hbox{\epsfxsize=3.30in \epsfbox{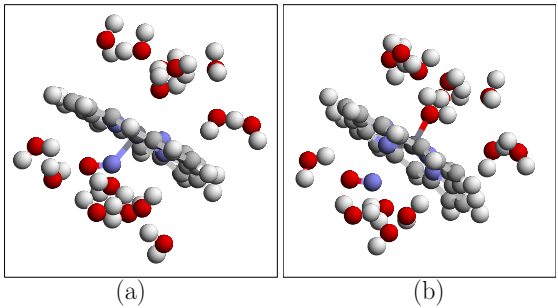}}}
\caption[]
{\label{fig6}
Representative snapshots of 
(a) initial and (b) final configurations of the Mn(III)P-NO complex in water
after 0.1~ps.  Only H$_2$O molecules close to the Mn ion are depicted.}
\end{figure}

Next, we consider Mn(III)P-NO in water.  Starting from an equilibrated
aqueous Mn(II)P-NO configuration (Fig.~1d), we remove an electron
from the simulation cell, and restart the DFT+U based AIMD trajectory with
the spin polarization fixed at $m=3/2$.  Within 0.1~ps,
the NO diffuses away from the Mn(III) ion, replaced by an
H$_2$O from the other side of the porphine plane
(Fig.~1e; Fig.~6).
Thus, DFT+U predicts that Mn(III)P fails to bind to NO in water,
in agreement with experiments.\cite{nitric1}
 
To examine PBE predictions for Mn(III)P-NO, we use a DFT+U
predicted Mn(III)P configuration in water, manually reposition
the NO molecule to yield a 180$^o$ Mn-N-O angle and a 1.8~\AA\,
Mn-N bond length, impose a low spin ($m=1/2$) spin state,
and restart the AIMD simulation using the PBE functional.
We find that Mn(III)P-NO complex is stable in water over the 
course of a 5~ps trajectory. 
While these AIMD runs are short, their qualitative results are completely
consistent with the relative stabiliy of Mn(III)P-NO and Mn(III)P-H$_2$O
based on gas phase binding energy considerations (Table~\ref{table1}).
They show that PBE predictions are in apparent disagreement with
experiments.\cite{nitric1,nitric2,note2}

\begin{table}\centering
\begin{tabular}{  l c c c c  } \hline
species & expt. & PBE & B3LYP & DFT+U \\ \hline
Mn(II)P & $s$=5/2 & no & yes & yes \\
Mn(III)P & $s$=2 & yes & yes & yes \\
Mn(II)P-NO & $s$=0 & yes & no  & no \\
Mn(III)P-NO & instability in water & no & NA  & yes \\ \hline
\end{tabular}
\caption[]
{\label{table2} \noindent
Brief summary of the successes and failures of different DFT methods.
Gas phase DFT and aqueous phase AIMD simulations that apply the
same functional predict the same stable spin states in all cases considered.
}
\end{table}
 
\subsection{Summary of MnP-NO interactions}
 
The apparent successes and failures of PBE and DFT+U methods
in the gas and liquid phases are summarized in Table~\ref{table2}.  
Recall that, in the absence of the NO ligand, PBE predicts an
incorrect, intermediate Mn(II)P ($s$=3/2) spin state while both B3LYP
and DFT+U parameterized with B3LYP spin splittings predict the
correct high spin ($s$=5/2) state.\cite{leung}  All three methods predict the
correct spin state for Mn(III)P (high spin, $s=2$).  The DFT+U method
correctly predicts that the Mn(III)P-NO is unstable in liquid water.  However,
the spin state of the predicted stable species in the gas phase is
inconsistent with the typical low-spin \{M-NO\}$^5$ structural parameters.
PBE predicts that Mn(III)P-NO is a strongly bound complex, in
apparent disagreement with aqueous phase experiments.
For the Mn(II)P-NO structure, PBE predicts the experimental structure
and spin state, while B3LYP and DFT+U do not.
 
Both the hybrid functional B3LYP and the DFT+U method favor the high-spin
states of first row transition metal ion, by increasing the exchange
interaction or screened couloumb interaction among $3d$ electrons.
To a first approximation, when Mn and NO are far apart,
enforcing $m=0$ on the Mn(II)P-NO complex requires 
equal and opposite local $s=1/2$ spin moments on
Mn and on the N atom of the NO molecule.
We conjecture that DFT+U cannot reproduce the stability of 
$m=0$ Mn(II)P-NO because Mn(II)P strongly favors the high spin
($m=5/2$) state, while the low spin ($m=1/2$) state is highly unfavorable.
In contrast, in PBE calculations the $m=1/2$ state is not extremely unfavorable
in energy compared to $m=3/2$, the stable Mn(II)P spin state erroneously
predicted by PBE.  Therefore, at least when Mn(II)P and NO are far apart,
PBE predictions more readily accommodates a $m=0$ spin state.
 
In summary, state-of-the-art DFT methods are as yet unable
to capture the delicate balance of competing effects and correlations
that determine the stable spin state of Mn(II)P, Mn(III)P, and
their nitric oxide complexes. NO ligation seems to be especially
challenging due to $\pi$-electron back bonding.\cite{feno,feno1}
The DFT+U method, with its most important parameter $U$ fitted
to the Mn(II)P high spin-intermediate spin splitting, should
be further improved to take into account such $\pi$-back bonding.\cite{note4}
These improvements may allow the more accurate prediction of binding
energies, which have not been addressed in recent DFT work
on porphyrin-nitric oxide complexes.\cite{feno,feno1}
 
%\begin{table}\centering
%\begin{tabular}{||l||r|r|r|r||} \hline
%$U$ (eV)       &  0.000 & 1.000 & 2.000 & 4.200 \\ \hline
%$\Delta E_1$ (eV) & -0.495 & & -0.343 & 0.214 \\
%$\Delta E_2$ (eV) &  0.548 &  0.666  & -0.324 & -0.432 \\
%\end{tabular}
%\caption[]
%{\label{table3} \noindent
%The effect of varying $U$ on DFT+U predictions
%of $E(s=3/2)$-$E(s=5/2)$ for Mn(II)P ($\Delta E_1$), and
%of $E(m=3/2)$-$E(m=1/2)$ for Mn(III)P-NO ($\Delta E_2$).
%Experimentally, both $\Delta E_1$ and $\Delta E_2$ are
%positive quantities.  $J=1$~eV in all calculations except
%for the $U=0$ case, which represent PBE calculations.
%}
%\end{table}
 
\section{Conclusions}
 
In this work, we have conducted {\it ab initio} molecular dynamics
simulations of the hydration environment of the Mn ion in Mn(II)P
and Mn(III)P dispersed in liquid water.  These are intended as
simple models of water-soluble manganese porphyrins.  The DFT+U
technique, parameterized with B3LYP spin splittings, successfully
predicts that high spin Mn(II)P and Mn(III)P are stable in liquid water.
Thus, this technique enables efficient molecular dynamics simulation of MnP
with the correct spin state in a condensed phase environment.  In contrast,
the B3LYP functional, with its long range exchange, is costly to apply
when using periodic boundary conditions.

The Mn ion in Mn(II)P is predicted to displaced off-center out of the
porphine plane
by an average of 0.48~\AA, and is ligated to a single H$_2$O molecule.
The Mn ion in Mn(III)P is on average co-planar within the porphine plane
and binds strongly to 2 H$_2$O molecules.  Water molecules ligated to
the Mn ion exhibit residence times of more than 15~ps.  Water only interacts
weakly with the nitrogen atoms on the porphine ring chelating the
Mn ion.  These predicted hydration structures might be potentially useful
for improved analysis of NMR relaxation data.\cite{expt2,expt3}
 
The application of DFT techniques to examine MnP-NO binding in both
gas and aqueous phases proves less successful than for MnP dispersed in
water.  Ideally, the DFT+U technique, fit to B3LYP spin splittings for
Mn(II)P and found to be successful for unligated MnP,
should be applicable to all ligated complexes as well.
Indeed, we find that DFT+U apparently succeeds in describing the
binding energy of the Mn(III)P-NO complex.
It predicts that the Mn(III)P-NO binding energy is substantially
smaller than that for Mn(III)P-H$_2$O.  AIMD based on the DFT+U technique
predicts a rapid dissociation of the Mn(III)P-NO in water, consistent
with the known instability of Mn(III)P-NO in aqueous phase
experiments.\cite{nitric1,nitric2}  Both
DFT+U and B3LYP predict significantly larger Mn(II)P-NO binding energies
than for Mn(III)P-NO.  This differential ability of Mn(II) and Mn(III)
porphyrins to ligate to NO is crucial for detecting NO in water and
distinguishing NO from HNO or NO$^-$ via electrochemical means.  In contrast,
PBE predicts large ($>$ 0.8~eV) Mn-NO binding energies for both Mn(II)P-NO
and Mn(III)P-NO, which would suggest, erroneously, that switching the
oxidation state of MnP cannot trap and then release NO.

%While no X-ray scattering data is available for NO-ligated Mn(III) porphyrin
%systems, 

We emphasize that B3LYP and DFT+U predictions qualiltatively agree 
with each other regarding Mn(II)P-NO binding,
despite the fact that the DFT+U technique used herein is parameterized using
B3LYP spin splitting for the bare Mn(II)P molecule.  However,
both these techniques predict incorrect spin states for the Mn(II)P-NO
complex.  On the other hand, the PBE functional predicts a Mn(II)P-NO
low spin spin state and molecular geometry that are in good agreement
with experiments.  
 
While it is clear that more fundamental improvements to DFT methods
need to be made, AIMD simulations based on the DFT+U technique
successfully predicts many experimentally observed ligation features of 
Mn porphyrins, including the observation
that complexes formed between Mn(III) porphyrins and nitric oxide
are unstable in liquid water.
 
\section*{Acknowledgements}
 
We thank John Shelnutt, Sameer Varma, and Prof.~Robert~Sharp for
useful discussions.  
Sandia is a multiprogram laboratory operated by
Sandia Corporation, a Lockheed Martin Company, for the U.S.~Department
of Energy's National Nuclear Security Administration
under contract DE-AC04-94AL8500.


\begin{references}

\bibitem{handbook}
Kadish,~K.~M.; Smith,~K.~M.; Guilard,~R. {\it The Porphyrin Handbook}
(Academic Press, San Diego, 2000).
 
\bibitem{nitric1}
M.~A.~Marti, S.~E.~Bari, D.~A.~Estrin, and F.~Doctorovich,
J.~Am.~Chem.~Soc. {\bf 127}, 4680 (2005).
 
\bibitem{nitric2}
J.~Oni, N.~Diab, I.~Radtke, and W.~Schuhmann, Electrochimica Acta
{\bf 48}, 3349 (2003).
 
\bibitem{nitric3}
I.~Spasojevic, I.~Batinic-Haberle, and I.~Fridovich, Nitric Oxide: Biol.
and Chem. {\bf 4}, 526 (2000). 
 
\bibitem{sol1}
B.~J.~Vesper, K.~Salaita, H.~Zong, C.~A.~Mirkin, A.~G.~M.~Barratt, and
B.~M.~Hoffman, J.~Am. Chem.~Soc. {\bf 126}, 16653 (2004).
                                                                                
\bibitem{sol2}
S.~Yoshimoto, J.~Inukai, A.~Tada, T.~Abe, T.~Morimoto, A.~Osuka,
H.~Furuta, and K.~Itaya, J.~Phys. Chem.~B {\bf 108}, 1948 (2004).
 
\bibitem{water}
Z.~Wang, C.~J.~Medfoth, and J.~A.~Shelnutt, J.~Am.~Chem.~Soc. {\bf 126}, 16720
(2004).

\bibitem{nitric_rev1}
W.~R.~Scheidt, K.~Hatano, G.~A.~Rupprecht, and P.~L.~Piciulo,
Inorg.~Chem. {\bf 18}, 292 (1979).
 
\bibitem{expt1}
N.~Schaefle and R.~R.~Sharp, J.~Chem.~Phys. {\bf 122}, 184501 (2005).
 
\bibitem{expt2}
J.~C.~Miller and R.~Sharp, J.~Phys.~Chem.~A {\bf  104}, 4889 (2000).
 
\bibitem{expt3}
L.~H.~Bryant, M.~W.~Hodges, and R.~G.~Bryant, Inorg.~Chem. {\bf 38},
1002 (1999).
 
\bibitem{expt4}
N.~Schaefle and R.~R.~Sharp, J.~Phys.~Chem.~A {\bf 109}, 3267 (2005).
 
\bibitem{dft_water}
D.~Asthagiri, L.~R.~Pratt, M.~E.~Paulaitis, and S.~B.~Rempe,
J.~Am.~Chem.~Soc. {\bf 126}, 1285 (2004).
 
\bibitem{threebody}
D.~R.~Nutt, M.~Karplus, and M.~Meuwly, J.~Phys.~Chem. B {\bf 109},
21118 (2005).
 
\bibitem{rode}
B.~M.~Rode, C.~F.~Schwenk, T.~S.~Hofer, and B.~R.~Randolf, Coord. Chem. Rev.
{\bf 249}, 2993 (2005).
 
\bibitem{reiher}
M.~Reiher, O.~Salomon, and B.~A.~Hess, Theor.~Chem.~Acc. 
{\bf 107}, 48 (2001); M.~Reiher,  Inorg.~Chem. {\bf 41}, 6928 (2002).
 
\bibitem{ghosh}
A.~Ghosh and P.~R.~Taylor, Curr. Opin Chem. Biol. {\bf 7}, 113 (2003).
 
\bibitem{b3lyp}
A.~D.~Becke, J.~Chem.~Phys. {\bf 98}, 5648 (1993);
C.~T.~Lee, W.~T.~Yang, and R.~G.~Parr, Phys.~Rev.~B 
{\bf 37}, 785 (1988).
 
\bibitem{leung}
K.~Leung, S.~B.~Rempe, P.~A.~Schultz, E.~M.~Sproviero, V.~S.~Batista,
M.~E.~Chandross, and C.~J.~Medforth, J.~Am.~Chem.~Soc. {\bf 128}, 3659 (2006).

\bibitem{pbe}
J.~P.~Perdew, K.~Burke, and M.~Ernzerhof, Phys.~Rev.~Lett.
{\bf 77}, 3865 (1996).

\bibitem{dimer}
Note that hybrid functionals such as B3LYP are not universally more
successful than non-hybrid ones like PBE.  A counter example is
the case of transition metal dimers. See N.~E.~Schultz, Y.~Zhao, D.~G.~Truhlar,
J.~Phys.~Chem.~A {\bf 109}, 4388 (2005) and references therein.
 
\bibitem{marz}
H.~J.~Kulik, M.~Cococcioni, D.~A.~Scherlis, and N.~Marzari,
{\tt http://arXiv.org/cond-matt/0608285} (2006)

\bibitem{note0}
The four negatively charged sulfonate groups in MnTSPP ensure that the
Mn ion is not ligated to anions in aqueous solutions, even if it is in 
the +3 oxidation state; hence we do not consider counter ions when
modeling Mn(III)P in liquid water.
 
\bibitem{ldau}
V.~I.~Anisimov, J.~Zaanen, and O.~K.~Andersen, Phys.~Rev.~B,
{\bf 44}, 943 (1991); A.~I.~Liechtenstein, A.~I.~Anisimov, and J.~Zaanen,
Phys.~Rev.~B {\bf 52}, 5467 (1995).
 
\bibitem{sprik}
Y.~Tateyama, J.~Blumberger, M.~Sprik, and I.~Tavernelli, J.~Chem.~Phys.
{\bf 122}, 234505 (2005).
 
\bibitem{nitric_rev}
G.~R.~A.~Wyllie and W.~R.~Scheidt, Chem. Rev. {\bf 102}, 1067 (2002),
and references therein.
 
\bibitem{nitric_rev2}
Z.~N.~Zahran, J.~Lee, S.~S.~Alguindigue, M.~A.~Khan,
and G.~B.~Richter-Addo, Dalton Trans. 44 (2004).
 
\bibitem{feno}
K.~M.~Vogel, P.~M.~Kozlowski, M.~Z.~Zgierski, and T.~G.~Spiro, 
J.~Am.~Chem.~Soc. {\bf 121}, 9915 (1999).
 
\bibitem{feno1}
T.~Wondimagegn and A.~Ghosh, J.~Am.~Chem.~Soc. {\bf 123}, 5680 (2001).
 
\bibitem{vaspldau}
O.~Bengone, M.~Alouani, P.~Bl\"{o}chl, and J.~Hugel,  Phys.~Rev.~B,
{\bf 62}, 16392 (2000).
 
\bibitem{vasp}
G.~Kresse and J.~Furthm\"{u}ller, Phys.~Rev.~B {\bf 54}, 11169 (1996);
Comput.~Mater.~Sci. {\bf 6}, 15 (1996).

\bibitem{paw}
P.~E.~Blochl, Phys.~Rev.~B, {\bf 50}, 17953 (1994); for VASP
implementation, see 
G.~Kresse and D.~Joubert Phys.~Rev.~B {\bf 59}, 1758 (1999).

\bibitem{overstructure}
D.~Asthagiri, L.~R.~Pratt, J.~D.~Kress, Phys.~Rev.~E {\bf 68}, 415051 (2003);
J.~C.~Grossman, E.~Schwegler, E.~W.~Draeger, F.~Gygi, G.~Galli,
J.~Chem.~Phys. {\bf 120}, 300 (2004);
J.~VandeVondele, F.~Mohamed, M.~Krack, J.~Hutter, M.~Sprik, M.~Parrinello,
J.~Chem.~Phys. {\bf 122}, 014515 (2005).
 
\bibitem{towhee}
M.~G.~Martin and A.~P.~Thompson, Fluid Phase Equil. {\bf 217}, 105 (2004).
 
\bibitem{spce}
H.~J.~C.~Berendsen, J.~R.~Gridera, and T.~P.~Straatsma,
J.~Phys.~Chem. {\bf 91}, 6269 (1987).
 
\bibitem{shelforce}
X.-Z.~Song, L.~Jaquinod, W.~Jentzen, D.~J.~Nurco, S.-L.~Jia,
R.~G.~Khoury, J.-G.~Ma, C.~J.~Medforth, K.~M.~Smith, and J.~A.~Shelnutt,
Inorg.~Chem.  {\bf 37}, 2009 (1998);
J.~A.~Shelnutt, C.~J.~Medforth, M.~D.~Berber,
K.~M.~Barkigia, and K.~M.~Smith, J.~Am.~Chem.~Soc. {\bf 113},
4077 (1991).
 
\bibitem{no}
Z.-W.~Zhou, B.~D.~Todd, K.~P.~Travis, and R.~J.~Sadus, J.~Chem.~Phys.
{\bf 123}, 054505 (2005).
 
\bibitem{g03}
M.~J.~Frisch, {\it et al.}, Gaussian 03 (Revision C.02), Gaussian Inc.,
(Wallingford, CT, 2004).
 
\bibitem{bery}
D.~Asthagiri and L.~R.~Pratt, Chem.~Phys.~Lett. {\bf 371}, 613 (2003).

\bibitem{expt_time}
Z.~Luz and R.~G.~Shulman, J.~Chem.~Phys. {\bf 43}, 3750 (1965);
R.~A.~Bernhein, T.~H.~Brown, H.~S.~Gutowsky, and D.~E.~Woessner, J. Chem.
Phys. {\bf 30}, 950 (1959).

\bibitem{marz1}
P.~H.-L. Sit and Ni.~Marzari, J.~Chem.~Phys. {\bf 122}, 204510 (2005).

\bibitem{galli2}
E.~Schwegler, J.~C.~Grossman, F.~Gygi, and G.~Galli, J.~Chem.~Phys.
{\bf 121}, 5400 (2004).

\bibitem{rigid}
K.~Leung and S.~B.~Rempe, Phys.~Chem.~Chem.~Phys. {\bf 8}, 2153 (2006).
 
\bibitem{zeropt}
These energies do not include zero point energy (ZPE) corrections.  We have not
computed ZPE for MnP, but we note that the FeP-NO vibration frequency
has been shown to be small, on the order of 550~cm$^{-1}$,\cite{feno} which
implies a ZPE of at most 2~kcal/mol.
 
\bibitem{mnst_expt1}
J.~F.~Kirner, C.~A.~Reed, W.~R.~Scheidt, J.~Am.~Chem.~Soc. {\bf 99},
1093 (1977).
                                                                                
\bibitem{mnst_expt2}
B.~Cheng, W.~R.~Scheidt, Acta Crystallogr. C. {\bf 52}, 361 (1996).

\bibitem{turner}
P.~Turner, M.~J.~Gunter, B.~W.~Skelton, A.~H.~White, and T.~W.~Hambley, 
J.~Chem.~Res.  {\bf 18}, 220 (1996).
 
\bibitem{note}
Indeed, our preliminary PBE based AIMD simulations, with the simulation
cell constrained to $m=5/2$, show that PBE predicts
a much higher tendency towards Mn(II) binding to two water molecules,
and the Mn(II) ion residing in the porphine plane.

\bibitem{hydrox}
See, e.g., K.~Leung, Biophys. Chem. (in press).
 
\bibitem{note1}
To make this point more precise, note that the
previous section shows that Mn(III)P binds to two H$_2$O in liquid
water.  Therefore we should compare the PBE energies of
Mn(III)P(NO)(H$_2$O) and Mn(III)P(H$_2$O)$_2$.  Table~\ref{table1}
indicates that that, in the gas phase, the former is more favorable
by 0.857~eV, or 19.75~kcal/mol.\cite{zeropt}  This is an underestimate of the
energetic preference for Mn(III)P(NO)(H$_2$O) in liquid water, because
replacing a NO ligand with H$_2$O from the bulk water region involves
breaking two hydrogen bonds and moving a hydrophobic species (NO) into
bulk water.  At T=300~K, one can readily estimate that the translational
entropy advantage of dissociating Mn(III)P-NO in a 1 $\mu$M NO solution is
$\sim 11$~kcal/mol.  This entropic gain is too small to offset the
binding energy cost for dissociation.  Thus, by any reasonable measure,
PBE erroneously predicts that the Mn(III)P(NO)
complex does not dissociates in liquid water.
 
\bibitem{note2}
The dynamics of dissociation can be extremely complex and may involve
spin-forbidden transitions.\cite{barrier}
Studying Mn(II)P-NO dissociation is not the
main objective of this work.
 
\bibitem{barrier}
J.~N.~Harvey, J.~Am.~Chem.~Soc. {\bf 122}, 12401 (2000).

\bibitem{note4}
We have not found values of $U$ within the DFT+U approach that yield
both a high spin ground state for Mn(II)P and a low spin ground
state for Mn(II)P-NO.  This suggests that dynamic electronic correlation
effects, not accounted for by the static DFT+U method, may be important
in Mn(II)P-NO.

\bibitem{epaps}
See EPAPS Document No. XXXX for supplemental information on
normal-coordinate structure decomposition (NSD) analysis of
several computed and experimental porphyrin structures.
This document may be retrieved via the EPAPS homepage
{\tt (http://www.aip.org/pubservs/epaps.html)} or from {\tt ftp.aip.org}
in the directory {\tt /epaps/}.  See the EPAPS homepage for more information.
 
\end{references}
\end{document}